\newif \ifclr   \clrfalse

\documentclass[11pt]{article}

\usepackage{fullpage}
\usepackage{graphicx}   
\usepackage{xcolor}      
\usepackage{amssymb,amsmath,mathabx}
\usepackage{booktabs,multirow} 

\newtheorem{theorem}{Theorem}

\newtheorem{proposition}[theorem]{Proposition}

\newtheorem{example}[theorem]{Example}

\newcommand\mypar[1]{\subsubsection{#1}}

\def\datum{\ensuremath{\mathbf{d}}}
\def\lab{\ensuremath{l}}
\def\data{\ensuremath{d}}
\newcommand{\st}{\ensuremath{^*}}

\def\Ialph{\ensuremath{\Sigma}}
\def\dataalph{\ensuremath{\mathbb{D}}}
\def\inputalph{\ensuremath{\mathbb{L}}}
\def\L{\ensuremath{{L}}}
\def\nat{\ensuremath{\mathbb{N}}}
\def\cl{\ensuremath{\mathsf{C}}}
\def\cs{\ensuremath{\mathsf{cs}}}

\newcommand{\dom}[1]{\ensuremath{dom(#1)}}
\newcommand{\domplus}[1]{\ensuremath{dom^+(#1)}}
\newcommand{\val}[2]{\ensuremath{val_{#1}(#2)}}
\def\leftmark{\ensuremath{\rhd}}
\def\rightmark{\ensuremath{\lhd}}

\def\RA{\ensuremath{\rm{RA}}}
\def\ra{\ensuremath{\mathcal{R}}}
\def\regval{\ensuremath{\tau}}
\def\initregval{\ensuremath{\regval_0}}

\def\lef{\rm{left}}
\def\sta{\rm{stay}}
\def\rig{\rm{right}}

\def\D{\ensuremath{\rm{D}}}
\def\N{\ensuremath{\rm{N}}}
\def\A{\ensuremath{\rm{A}}}

\def\PA{\ensuremath{\rm{PA}}}
\def\pa{\ensuremath{\mathcal{P}}}

\def\lcp{\text{\rm{lift-current-pebble}}}
\def\pnp{\text{\rm{place-new-pebble}}}

\def\P{\ensuremath{P}}
\def\V{\ensuremath{V}}

\def\DA{\ensuremath{\rm{DA}}}
\def\da{\ensuremath{\mathcal{D}}}
\def\daA{\ensuremath{\mathcal{T}}}
\def\daB{\ensuremath{\mathcal{A}}}

\def\CMA{\ensuremath{\rm{CMA}}}
\def\cma{\ensuremath{\mathcal{C}}}
\def\cmastates{\ensuremath{Q}}

\def\cmatrans{\ensuremath{\delta}}

\def\cmafinloc{\ensuremath{F_L}}
\def\cmafinglo{\ensuremath{F_G}}
\def\cmaf{\ensuremath{\mathsf{f}}}

\def\equal{\ensuremath{\sim}}
\def\FO{\text{\rm{FO}}}
\def\MSO{\text{\rm{MSO}}}
\def\EMSO{\text{\rm{EMSO}}}
\def\foelp{\ensuremath{\text{\rm{FO}(}\hspace*{-0.1cm}\equal,<,\text{\rm{+1)}}}}
\def\foep{\ensuremath{\text{\rm{FO}(}\hspace*{-0.1cm}\equal,\text{\rm{+1)}}}}
\def\fotelp{\ensuremath{\text{\rm{FO}}^2\text{\rm{(}}\hspace*{-0.1cm}\equal,<,\text{\rm{+}}1\text{\rm{)}}}}
\newcommand\fokelp[1]{\ensuremath{\text{\rm{FO}}^#1\text{\rm{(}}\hspace*{-0.1cm}\equal,<,\text{\rm{+}}1\text{\rm{)}}}}
\newcommand\fokel[1]{\ensuremath{\text{\rm{FO}}^#1\text{\rm{(}}\hspace*{-0.1cm}\equal,<\text{\rm{)}}}}
\def\fotep{\ensuremath{\text{\rm{FO}}^2\text{\rm{(}}\hspace*{-0.1cm}\equal,\text{\rm{+1)}}}}
\def\fotel{\ensuremath{\text{\rm{FO}}^2\text{\rm{(}}\hspace*{-0.1cm}\equal,<\text{\rm{)}}}}
\def\fothreeelp{\ensuremath{\text{\rm{FO}}^3\text{\rm{(}}\hspace*{-0.1cm}\equal,<,\text{\rm{+}}1\text{\rm{)}}}}
\def\foelw{\ensuremath{\text{\rm{FO}}\text{\rm{(}}\hspace*{-0.1cm}\equal,<,\text{\rm{+}}\omega\text{\rm{)}}}}
\def\fotelt{\ensuremath{\text{\rm{FO}}^2\text{\rm{(}}\hspace*{-0.1cm}\equal,<,\text{\rm{+}}1,\oplus 1\text{\rm{)}}}}
\def\fotelw{\ensuremath{\text{\rm{FO}}^2\text{\rm{(}}\hspace*{-0.1cm}\equal,<,\text{\rm{+}}\omega\text{\rm{)}}}}
\def\msoelp{\ensuremath{\text{\rm{MSO}(}\hspace*{-0.1cm}\equal,<,\text{\rm{+1)}}}}

\newcommand\msokelp[1]{\ensuremath{\text{\rm{MSO}}^#1\text{\rm{(}}\hspace*{-0.1cm}\equal,<,\text{\rm{+1)}}}}
\def\msoep{\ensuremath{\text{\rm{MSO}(}\hspace*{-0.1cm}\equal,\text{\rm{+1)}}}}
\def\emsoelp{\ensuremath{\text{\rm{EMSO}}\text{\rm{(}}\hspace*{-0.1cm}\equal,<,\text{\rm{+}}1\text{\rm{)}}}}
\def\emsotelp{\ensuremath{\text{\rm{EMSO}}^2\text{\rm{(}}\hspace*{-0.1cm}\equal,<,\text{\rm{+}}1\text{\rm{)}}}}
\def\emsotelw{\ensuremath{\text{\rm{EMSO}}^2\text{\rm{(}}\hspace*{-0.1cm}\equal,<,\text{\rm{+}}\omega\text{\rm{)}}}}
\def\emsotelt{\ensuremath{\text{\rm{EMSO}}^2\text{\rm{(}}\hspace*{-0.1cm}\equal,<,\text{\rm{+}}1,\oplus 1\text{\rm{)}}}}
\def\emsoteltw{\ensuremath{\text{\rm{EMSO}}^2\text{\rm{(}}\hspace*{-0.1cm}\equal,<,\text{\rm{+}}\omega,\oplus 1\text{\rm{)}}}}
\newcommand\emsokelp[1]{\ensuremath{\text{\rm{EMSO}}^#1\text{\rm{(}}\hspace*{-0.1cm}\equal,<,\text{\rm{+1)}}}}
\newcommand\emsokelt[1]{\ensuremath{\text{\rm{EMSO}}^#1\text{\rm{(}}\hspace*{-0.1cm}\equal,<,\text{\rm{+}}1,\oplus\text{\rm{1)}}}}

\def\tfotelp{\ensuremath{\text{\rm{FO}}^2\text{\rm{(}}\hspace*{-0.1cm}\equal,E_{\rightarrow},E_{\downarrow},E_{\rightarrow}\st,E_{\downarrow}\st\text{\rm{)}}}}
\def\tfotep{\ensuremath{\text{\rm{FO}}^2\text{\rm{(}}\hspace*{-0.1cm}\equal,E_{\rightarrow},E_{\downarrow}\text{\rm{)}}}}
\def\temsotep{\ensuremath{\text{\rm{EMSO}}^2\text{\rm{(}}\hspace*{-0.1cm}\equal,E_{\rightarrow},E_{\downarrow}\text{\rm{)}}}}

\def\outputalph{\ensuremath{\Gamma}}

\newcommand{\loris}[1]{{{\color{blue}{\ifclr{L: #1}\else \empty \fi}}}}
\newcommand{\comp}[1]{\textsc{#1}}

\title{In the Maze of Data Languages}
\author{Loris D'Antoni\\
University of Pennsylvania
}
\date{}
\begin{document}

\maketitle

\begin{abstract}
In data languages the positions of strings and trees carry 
a label from a finite alphabet and a
data value from an infinite alphabet.
Extensions of automata and logics over finite alphabets have been defined 
to recognize data languages, both in the string and tree cases.
In this paper
we describe and compare the complexity and expressiveness of such models
to understand which ones are better candidates as \emph{regular} models.
\end{abstract}

\section{Introduction}
\loris{need to add emph around}

The presence of values ranging over an infinite domain in everyday programming has led to the study of languages over infinite alphabets. 
In areas like
program verification \cite{Henzinger92timedtransition,SSTAlur,Bouajjani2002,Demri2009} and XML processing and validation \cite{Bojanczyk2006tree},
\loris{add more applications take some from regularity paper}
the finite alphabet abstraction is too restrictive 
to perform useful analysis. 

In program verification, for example, 
unbound data and properties about infinite domains (like natural numbers) are commonly used.
Unfortunately, when dealing with infinite domains most problems are undecidable. 
However, not everything is lost. 
In fact, some problems become decidable when restricting the set of allowed operations and/or the number of variables ranging over an infinite domain \cite{Bouajjani2002}.

In terms of XML processing,
the problem of representing infinite domains occurs in several settings.
Although the tree structures in XML range over a finite alphabet, the values that can be stored in the XML attributes do not \cite{Alon2001}.
Most languages model XML documents with labelled, ordered, unranked trees, where the labels
are from a finite alphabet, while the attribute values are usually ignored. 
Clearly, the current approach is too limiting, especially if we want to ask questions
about the values contained in the attributes, such as, \emph{`Does an XML specification accept trees where two nodes contain the same attribute values?'}.
Again, crossing the line of undecidability is quite easy. 
However, several restrictions with interesting decidable properties and good expressiveness have been proposed.

One of the most studied and successful restriction is the one where
\emph{data values} from an infinite domain can be attached to the labels over a finite domain that appears in our program, string, tree, etc \cite{Bojanczyk2006word,Bojanczyk2006tree}.
In particular, in a \emph{data string} each position in the string carries a label from a finite alphabet and a data value from an infinite domain.
A \emph{string data language} is a language (set) of data strings.
With this framework set, the following question arises:
\emph{``What is a regular data language?''} 
Several people have tried answering this question with new automata models and new logics over data strings.

When we hear the term \emph{regular language}, we immediately think of finite automata and strings over a finite alphabet.
Regular string languages gave rise to a very important concept in computer science: \emph{regularity of languages}.
But how do we extend this concept to string data languages?
Before attempting to answer this, let's ask ourselves a couple of other questions to better understand the notion
of regularity. 
Why are regular string languages so popular? What made them
an appealing model? 

Regular string languages have a 
good trade-off between expressiveness and decidability. 
They are also closed under a broad set of operations, and most of the interesting
decision problems can be solved in polynomial time. Every regular language can also be expressed as a formula in monadic second-order logic (\MSO) over strings \cite{langautomlogic}.

Then we might ask a more general question: \emph{``What makes a language regular?''}
All the previously mentioned properties make up a good answer, but 
the notion of language regularity can be generalized further.
A \emph{regular language} should have the following properties.
\begin{description}
\item[Expressiveness vs Decidability.] They should capture the largest class of languages where properties such as
	emptiness, equivalence, inclusion, etc are decidable.
\item[Efficiency.] Testing whether a given element belongs to a regular language should be solvable in linear time.
\item[Closure properties.] They should be closed under operations such as intersection, complement, reverse, etc.
\item[Robustness.] The class of languages they define should have many characterizations, in particular an equivalent
representation in logic.
\end{description}

All these properties hold for \emph{regular string languages}
and the notion of regularity has been successfully extended to other kinds of structures (languages), including infinite strings and finite or infinite, ranked or unranked trees. 
Many equivalent models and logics for representing regular languages over these structures have been proposed and
the properties listed above are satisfied by these models. The notion of regularity has also been extended to transductions (i.e. transformations from input strings/trees to output strings/trees). 
In all these cases the generalizations have been quite straightforward, in the sense that they simply used natural extensions of finite automata and monadic second-order logic.
Unfortunately, for data languages, the story is not as pleasant. 
When we try to extend the standard models to data languages,
even the simplest properties become undecidable. 

Since simple extensions did not work, new models have been proposed to deal with data languages.
To define a language, there are mainly two approaches: the automata approach and the logic approach. 
Both the approaches have been explored in the case of data languages.
Even though the automata approach is closer to the notion of computation, the logic approach is easier to reason about.
Usually regularity occurs when a decidable logic meets an automata model. 
As we mentioned earlier,
the equivalence of $\MSO$ over strings and finite automata
is a notable example. 

We present the major models proposed for data languages and compare their properties.
We will not present proofs, which can be found in the references.
For \emph{string data languages} several models have been proposed in the last decade.
In \cite{NevenFSM} Neven et al. provide a complete analysis on expressiveness and decidability for two automata models:
\emph{register automata} and \emph{pebble automata}. 
Register automata are finite automata equipped with a finite set of registers that can hold data values which can be tested for equality.
Pebble automata are finite automata equipped with a finite set of pebbles that can hold pointers to positions in the string.
Pebbles must be accessed with a stack discipline.
Although most of the results in \cite{NevenFSM} are negative, their 
paper renewed interest in the study of languages over infinite alphabets. 
In the paper they also compare several variants of these models and describe how they behave with respect to 
logics over data. Although this kind of comparison is standard when talking about languages, it does not
work as well in the case of data languages. 
In fact, even in a simple logic like first-order logic (\FO) with equality test and 
ordering over string positions, the satisfiability of a formula is already an undecidable problem.

A better idea is proposed in \cite{Bojanczyk2006word}, where decidability
for first-order logic over data is studied in detail. 
If formulae are allowed to have at most two variables,
checking the satisfiability of a first-order formula over data strings is proven to be decidable. 
Moreover, even with three variables the problem already is undecidable. 
The decidability results are then extended to more powerful logics with the same limitation on the number of
variables. Finally, they propose a new model called \emph{data automata} which
they prove to be equivalent to one of the decidable logics. Data automata inspect the input in two passes using
a transducer over the labels of the string and an NFA over chunks of the output of the transducer.

Data automata are limited by their difficult usage.
In \cite{Björklund07}, \emph{class memory automata} are introduced as an equivalent variant of data automata. 
Class memory automata perform operations similar to data automata but in a single pass.
The new model is also compared to previous models, such as register automata, and several decision problems are studied.
To mention a different approach, an algebraic characterization of some classes of data languages is presented in \cite{Bouyer2003}.
In section \ref{stringdef}, we present several of these models over string data languages and summarize their expressiveness and complexity results.

Just as there are models for data strings, there are proposed models for \emph{tree data languages} \cite{Cheng1998,Kaminski2008}.
While for the string case the area is quite advanced, tree data languages are still not well understood.
For languages over finite alphabets, natural extensions of string models have led to positive results with the addition 
of a hierarchical structure.
As an example, all the decidability results for $\FO$ and $\MSO$ over string languages
still hold for tree languages and, with small modifications, even for transducers.
However, in the case of data languages, this approach does not work. 
When we extend two-variable first-order logic to work on data trees, 
checking whether a given formula is satisfiable becomes a difficult problem.
Moreover models like register and pebble automata do not extend naturally to trees. 

The only positive result, so far, is that satisfiability  
for a small fragment of two-variable $\FO$ over trees is decidable \cite{Bojanczyk2006tree} (section \ref{treedef}).
A variant of class memory automata that works over trees has been proposed in \cite{Bojanczyk2012}. 
However most of 
the results are negative because the model is too expressive.
Otherwise, the area of data tree languages is still at an embryonic stage.

There has also been some progress in the area of transducers over infinite alphabets \cite{Veanes2012}.
However, no good model (even in logic) has been proposed yet for data transducers. 
In \cite{Veanes2012}, a new approach for representing languages and transductions over infinite alphabets is presented. 
The approach separates the theory of the data from the theory of the automata. 
The new models are called \emph{symbolic automata} and \emph{symbolic transducers}. 
In this setting, expressiveness is limited, but the set of operations that can be performed on the input data is not restricted to equality. 
Instead, the input (and output in the case of transducers)
alphabet is equipped with a decidable theory over it. 
If the theory is decidable in polynomial time, several problems such as checking equivalence, computing pre-images, etc are also
decidable in polynomial time.

At the end of the paper (section \ref{discussion}) we give a summary of the results and explore future research directions and open problems
in the field of data languages.

\section{String Data Languages}
\label{stringdef}

In this section we present several models of automata and logics over data strings. 
Their expressiveness, closure properties, and their decision properties are then compared.
Based on this analysis, we discuss which ones are better candidates to be regular models.

\subsection{Basic Definitions}
\label{sec:basicdef}
A \emph{datum} $\datum$ is a pair $(\lab,\data)$
where $\lab$ is a label from a finite alphabet $\inputalph$ and $\data$ is an element (data) of an
infinite alphabet $\dataalph$ (the data alphabet). The infinite alphabet $\inputalph\times\dataalph$ is denoted with $\Ialph$.
A \emph{data string} is a string over an infinite alphabet $\Ialph$. 
Formally a data string $w$ is a sequence $(\lab_1,\data_1)\ldots(\lab_n,\data_n)$, where each $(\lab_i,\data_i)\in\Ialph$ and $n\geq 0$.
The length $|w|$ of $w$ is $n$ and we write $\dom{w}$ for the set $\{1,\ldots,n\}$ of positions in $w$. 
For $i\in\dom{w}$, we also write $\val{w}{i}$ for $(\lab_i,\data_i)$. 
A \emph{string data language} is a set of data strings\footnote{In \cite{NevenFSM} data strings are over $\dataalph$. Even though
this model might seem simpler, it does not naturally extends string languages and it does not have a clear counter part in
practical applications like XML processing. Moreover, no properties about the underlying label language can be expressed.}.

Without loss of generality, since we will be dealing with two-way automata, we delimit input strings with two special symbols,
$(\leftmark,\leftmark), (\rightmark,\rightmark)\notin \Ialph$ for the left and the right ends of the the string. 
Notice that $\leftmark$ and $\rightmark$ do not appear in $\inputalph\cup\dataalph$. 
Throughout the paper we often just write $\leftmark$ for$(\leftmark,\leftmark)$ (same for $\rightmark$).
Given a string $w=\leftmark v\rightmark$, where $v\in\Ialph\st$, the positions of $\leftmark$ and $\rightmark$ will be $0$ and $|v|+1$ 
respectively. We write $\domplus{w}$ for the extended set $\{0,\ldots,|v|+1\}$ of positions.
We extend the $\val{w}{\cdot}$ function in the natural way: $\val{w}{0}=\leftmark$ and $\val{w}{|v|+1}=\rightmark$.
For consistency with the pair notation, $\leftmark$ and $\rightmark$ will actually be considered of the form $(\leftmark,\leftmark)$ and $(\rightmark,\rightmark)$.

\subsection{Models}
In the following we give semi-formal definitions of several models of automata and logics for data languages.
The reader can find the formal definitions in the references.
This list is not complete, but it tries to capture the historically interesting models in the attempt to define
regular data languages. 
\mypar{Register Automata}
\label{sec:regaut}
A \emph{register automaton} (\RA) \cite{NevenFSM,Demri2009} is a finite state machine equipped with a finite set of registers. 
Registers hold values from an infinite domain $\dataalph$. 
When processing an input symbol $(\lab,\data)\in\Ialph$, a $\RA$ compares $\data$ (checks for equality) with the values in the registers. 
Based on the comparison it can store the current symbol in one of the registers. The only possible operations that can be performed on a register
are: 1) checking equality with the data in the current input, and 2) updating the register to the current data. When reading the current symbol
the head of the automaton can move left, right or stay in the same position and the current state can be updated.

In a $\RA$ transitions are of the following forms:
\begin{itemize}
	\item $(i,q,\lab) \rightarrow (q',m)$, or 
	\item $(q,\lab) \rightarrow (q',i,m)$.
\end{itemize}
$q$ and $q'$ are states, $i$ refers to the $ith$ register and $m\in\{\sta, \lef, \rig\}$ is the direction in which the head of the automaton moves when performing the transition.

Informally, at every point, 
a transition of the form $(i,q,\lab)\rightarrow (q',m)$ apply if the current data value is stored in the $ith$ register and 
a transition of the form $(q,\lab)\rightarrow (q',i,m)$ apply if the current data value is not stored in any register. In this case the current data value is stored in register $i$.

An automaton starts processing a data string with an initial register assignment that assigns some starting data values to the registers.
To allow the automata to recognize the symbols $\leftmark,\rightmark$, they are always contained in the initial register assignment.
Configurations of the $\RA$ will be triplets containing the current position of the string, the current register assignment and the current state.
A data string $w$ is accepted 
if the automaton reaches a configuration in which the current state is a final state.

A register automaton is \emph{deterministic} if in any configuration at most one transition applies.
An automaton is \emph{one-way} if there are no left transitions.

\begin{example}
\label{ex:raconsec}
Let $\L_0$ be the following language of data strings. A data string $w$ is in $\L_0$ iff there exists two positions
$j$ and $j+1$ in $w$, such that $j$ and $j+1$ contain the same data value.
We define a one-way, deterministic, one-register $\RA$ $\ra$ that recognizes $\L_0$.
$\ra$ is a one-register automaton and is defined as follows:
\begin{itemize}
\item the set of states is $\{q_0,q_1,q_f\}$,  and $q_f$ is the only final state, 
\item the register $1$ contains the left-mark $\initregval(1)=\leftmark$, and 
\item $\ra$ has the following transitions: for every $\lab\in\inputalph$
 	\begin{enumerate}
		\item $(1,q_0,\leftmark)\rightarrow (q_1,\rig)$;
		\item $(1,q_1,\lab)\rightarrow (q_f,\sta)$;
		\item $(q_1,\lab)\rightarrow (q_1,1,\rig)$.
	\end{enumerate}
\end{itemize}
Notice that when the head reaches $\rightmark$, no transition applies since $\rightmark\notin\inputalph$.
At every point the register contains the data value $\data$ of the previous positions. If the current position also contains
$\data$ (2), $\ra$ goes to the final state. If the current position holds a value $\data'$ different from $\data$ (3), the head moves right
and the register is updated to $\data'$.
\end{example}

An \emph{alternating register automaton} $\ra$ keeps a set of so called \emph{universal states}.
Usually, when an automaton is in a configuration with state $q$, and is reading a data string $w$, we say that $w$ is accepted from the current configuration there exists a sequence of transitions that leads the automaton to a final states.
In the case of alternating automata this definition is slightly different.
When the alternating \RA, while reading a data string $w$, is in a configuration with a universal state $q$, and a set of transitions applies,
we say that $w$ is accepted from the current configuration if it is also accepted from \emph{all} the configurations reached applying one of
the possible transitions. 
If the current configuration is in a state $q$ that is not universal, and a set of transitions applies,
we say that $w$ is accepted from the current configuration if it is also accepted from \emph{one of the} the configurations reached applying one of the possible transitions. 
Whenever $q$ is a final state, $w$ is accepted from the current configuration.


We introduce some notation to indicate the different kinds of automata and the class of languages they define.
We use the notation wC-$\RA$ to refer to the different classes of $\RA$, where $w\in\{1,2\}$ stands for 1 or 2-way, and
$C\in\{\D,\N,\A\}$ stands for deterministic, non-deterministic and alternating respectively.
We use the same names to express a formalism and its corresponding class of definable languages.
For example, we write {\rm{1}}\D-$\RA$ $\subseteq$ {\rm{2}}\D-$\RA$ to say that the class of languages accepted by a {\rm{1}}\D-$\RA$ is 
included in the class of languages accepted by a {\rm{2}}\D-$\RA$.

\mypar{Pebble Automata}
\label{sec:pebaut}
\loris{re-check if definitions are too similar to papers}
\emph{Pebble automata} (PA) \cite{NevenFSM} are finite state automata with a finite ordered set $\{1,\ldots,k\}$ of pebbles where every pebble points to a position in the string.
Besides ordinary actions, when reading an input, a pebble automaton can add a new pebble to the string or lift the current pebble from the string.
These two actions must respect a stack discipline; we can lift the $i$\emph{th} pebble only if the $(i+1)$\emph{th} pebble is not present on the string, and
we can add the $i$\emph{th} pebble only if all the pebbles $\{1,\ldots,i-1\}$ are present on the string.
The highest pebble present on the string acts as the head of the automaton.
A transition depends on the current state, the current set of pebbles placed on the string, and the highest pebble placed on the string.
With a transition the automaton can change the current state and update the pebbles by moving the head left or right, lifting the current pebble, or adding a new pebble.
When the $i$\emph{th} pebble is removed, the $(i-1)$\emph{th} pebble, becomes the new head.

Transitions are of the form $(i, q, \lab, \P, \V) \rightarrow (q',m)$ where $i$ represents the $ith$ pebble, $q$ is the current state,
$\lab$ the label being read, $\P,\V\subseteq \{1,\ldots,i-1\}$ are subsets of pebbles, and
$$m\in\{\sta,\lef,\rig,\pnp,\lcp\}.$$

Informally, a transition $(i, q, \lab, \P, \V) \rightarrow (q',m)$ applies if pebble $i$ is the current head, 
$i$ is placed on the position $j$ in the string containing the value $(\lab,\data)\in \Ialph$, $q$ is the current state, $\V$ is the set of pebbles whose position contains the data value $\data$, and $\P$ is the set of pebbles whose position is $j$.
When the transition applies the current state is updated to $q'$ and the head of the automaton (the highest pebble) moves based
on the value of $m$. If $m=\pnp$ the $(i+1)th$ pebble is placed in the same positions of the $ith$ pebble. If $m=\lcp$ the $ith$
pebble is lifted and the $(i-1)th$ pebble becomes the head of the automaton. In the cases of $\sta, \lef, \rig$, the position of the pebble
remains the same, gets decremented, gets incremented respectively.


The notions of \emph{accepting language}, \emph{deterministic}, \emph{two-way} and \emph{alternating} are defined in the same way as before.
A possible alternative to the previous model places new pebbles at position $0$ in the string instead of at the position of the most recent
pebble.
This alternative is not relevant in the case of two-way pebble automata, but is crucial in the one-way case.
The model we presented before is referred to as \emph{weak} pebble automata, while the alternative one is called
\emph{strong} pebble automata. 

We use the notation wC-$\PA$ to refer to the different variants of $\PA$, where $w\in\{1,2\}$, and
$C\in\{\D,\N,\A\}$ have the same meanings as for $\RA$.
To indicate \emph{weak} and \emph{strong}, when it makes a difference, we prefix the notation with W and S, respectively.
Again, we use the same names to express a formalism and the corresponding class of definable languages.

\begin{example}
We consider the language $\L_0$ of Example \ref{ex:raconsec}.
We define a one-way, deterministic, two-pebble $\PA$ $\pa$ that recognizes $\L_0$.
$\pa$ is defined as follows:
\begin{itemize}
\item the set of states is $\{q_\leftmark,q_1,q_2,q_r,q_f\}$, with $q_f$ the only final state, 
\item the first pebble is on the first position in the data string,
\item $\ra$ has the following transitions; for every $\lab\in\inputalph$
 	\begin{enumerate}
		\item $(1,q_1,\lab,\emptyset,\emptyset)\rightarrow (q_r,\pnp)$;
		\item $(1,q_r,\lab,\emptyset,\emptyset)\rightarrow (q_1,\rig)$;
		\item $(2,q_r,\lab,\{1\},\{1\})\rightarrow (q_2,\rig)$;
		\item $(2,q_2,\lab,\emptyset,\{1\})\rightarrow (q_f,\sta)$;
		\item $(2,q_2,\lab,\emptyset,\emptyset)\rightarrow (q_r,\lcp)$.
	\end{enumerate}
\end{itemize}
The automaton uses pebbles 1 and 2 to perform the comparison on adjacent positions.
$\pa$ is in state $q_1$ if only pebble 1 is present and $\pa$ is going to start a sequence of transitions that will check if the data in $i$ (the position currently in pebble 1) is the same as that in position $(i+1)$.
Starting from state $q_1$ the automaton places a new pebble (1) and moves the head of the current pebble (i.e. pebble 2) to position
$(i+1)$ (3). Now that the positions $i$ and $(i+1)$ are stored in pebbles 1 and 2, $\pa$ checks if the value in position $(i+1)$ is the same as the value in position $i$ (4-5); if the value is the same (4), $\pa$ goes to an accepting state;
otherwise (5), $\pa$ lifts the current pebble and moves the head to position $(i+1)$ (2) to restart the sequence of transitions.
\end{example}

\mypar{Data Automata}
\label{sec:dataaut}
Data automata \cite{Bojanczyk2006word,Björklund07} were introduced as an extension of
{\rm{1}}\N-$\RA$ with better connections to logic.
$\RA$s and $\PA$s are only able to express properties about adjacent positions, but not about positions in the same class (i.e. positions containing the same data value).
A \emph{class} of $w$ is a maximal subset $\cl$ of $\dom{w}$, such that all the positions in $\cl$ hold the same data value.
We consider $\dom{w}$ instead of $\domplus{w}$ because the automaton is one way.
Given a data string $w=(\lab_1,\data_1)\ldots(\lab_n,\data_n)$ and a class $\cl=\{x_1,\ldots,x_k\}$ of $w$ where 
$x_1<\ldots<x_k$, we define the \emph{class string} of $\cl$ (we write $\cs(\cl)$) to be the string $\lab_{x_1}\ldots\lab_{x_k}$.

A data automaton runs on data strings in two passes. 
During the first pass a letter-to-letter string transducer
simply changes the label of each position without affecting the data. 
The second pass is over the result of this translation:
an NFA runs on each
sequence of letters having the same data value (class strings).

A letter-to-letter string transducer, from $\outputalph_1$ to $\outputalph_2$, is a finite state machine that, when
reading a input symbol from $\outputalph_1$, outputs a symbol over a finite output alphabet $\outputalph_2$.
Formally a non-deterministic letter-to-letter string transducer $\daA$ from $\outputalph_1$ to $\outputalph_2$ is a tuple $(Q,q_0,F,T)$
where $Q$ is a set of states, $q_0\in Q$ is an initial state, $F\subseteq Q$ is a set of final states, and 
$T\subseteq Q\times\outputalph_1\times Q \times\outputalph_2$
is a set of transitions. The semantics is the one expected: a transition $(q,a,q',b)$ applies if $\daA$ is in state $q$, and the next input symbol is $a$; if the transition is taken, $\daA$ goes to state $q'$ and outputs $b$.  
A run of $\daA$ on a string $w$ is accepting if when reading $w$ starting in $q_0$ the transducer ends in a state $q\in F$.
Given a string $s=a_1\ldots a_n\in\outputalph_1\st$, we use $\daA(s)$ to denote the set of possible outputs of $\daA$
on $s$.

A \emph{data automaton} $\da$ over the alphabet $\Ialph=\inputalph\times\dataalph$ is then a triple $(\daA, \daB, \outputalph)$ where $\daA$ is a non-deterministic letter-to-letter string transducer (called the \emph{base automaton})
from $\inputalph$ to $\outputalph$, and $\daB$ is an NFA (called the \emph{class automaton}) over $\outputalph$.
In the following we usually omit $\outputalph$ when it is clear from the context.

A data automaton $\da$ accepts a data string $w = (\lab_1,\data_1) \ldots(\lab_n,\data_n)$, iff
there exists a string $b_1\dots b_n$ in $\daA(\lab_1\ldots\lab_n)$, such that, for
every class $\cl=x_1,\ldots,x_k$ of $w$, the class automaton $\daB$ accepts
the string $b_{x_1}\ldots b_{x_k}$.

\begin{example}
\label{ex:ldif}
Consider the language of data strings $\L_{dif}$.
A data string $w$ belongs to $\L_{dif}$ if all the positions of $w$ contain different data values.
We define a data automaton $\da=(\daA,\daB)$ that recognizes $\L_{dif}$.
$\daA$ simply translates every symbol in $\inputalph$ to the symbol $0$ ($\outputalph=\{0\}$).
The class automaton $\daB$ checks that every class string is equivalent to $0$. We omit the formal definition. Notice
however that this language cannot be defined by a register automaton \cite{Segoufin2006}.

We can also define a data automaton $\da'$ that accepts the complement of  $\L_{dif}$.
A data string $w$ belongs to $\widebar{\L_{dif}}$ if there exists two distinct positions with the same data value.
$\da'=(\daA,\daB)$ works in the following way.
$\daA$ non-deterministically selects two positions and outputs $1$ when reading them and $0$ when reading all the other positions ($\outputalph=\{0,1\}$).
The class automaton $\daB$ checks that every class string contains either no $1$s or exactly two $1$s.
We omit the formal definition.
\end{example}

\mypar{Class-Memory Automata}
\label{sec:cma}
The main limitation of data automata is that they are hard to use in practice. Moreover, no notion
of determinism is defined for them.
In \cite{Björklund07}  \emph{class-memory automata} (\CMA) were introduced as a simplification of data automata that overcomes these
problems.

A $\CMA$ $\cma$ runs once, left to right, over the input data string. As usual it has a set of states and an initial state.
A $\CMA$ has two kinds of accepting states $\cmafinloc$ and $\cmafinglo$. 
They are respectively called local and global accepting states.
Transitions are of the form $(q, \lab, \bot) \rightarrow q''$ or $(q, \lab, q') \rightarrow q''$.
Informally the first kind of transition applies when $\cma$ reads a data value $\data$ for the first time and it is are in state $q$.
The second kind of transition applies when $\cma$ is in state $q$, the current symbol is of the form $(\lab,\data)$, and 
the previous time $\cma$ read a symbol with data value $\data$ the transition moved to state $q'$.
This mechanisms allows the $\CMA$ to process class strings.
A data string is accepted if the final state reached by $\cma$ is a globally accepting state and for every data value $\data$ in the string,
the last time a symbol with data value $\data$ was processed by $\cma$ the transition moved to a locally accepting state.

A $\CMA$ is deterministic if it doesn't have two transitions with the same left-hand side. 
We denote them with $\D\CMA$s.
\begin{example}
We construct a $\CMA$ $\cma$ that recognizes $\L_{dif}$, the language presented in Example \ref{ex:ldif}.
The set of states $\cmastates$ is $\{q_0,q_1,q_2\}$, with $q_0$ initial state.
The set of globally accepting states is $\cmafinglo=\cmastates$. This is due to the fact that we do not care about labels
since the language does not describe any global property.
The set of locally accepting states $\cmafinloc$ will be the one representing whether a particular datum has been repeated
or not. Informally, at every point in a run, for every data value $\data$, $\cmaf(\data)$ will contain: 
1) $\bot$, if $\data$ has not appeared in the string yet,
2) $q_1$, if $\data$ has appeared exactly once,
3) $q_2$, if $\data$ has appeared more than once.
$\cmafinloc$ will then be the set $\{q_1\}$.
The transition relation is now easy to define.
For every $\lab\in\inputalph$, for every $q\in\cmastates$, we will have:
\begin{itemize}
\item $\cmatrans(q,\lab,\bot)\rightarrow q_1$;
\item $\cmatrans(q,\lab,q_1)\rightarrow q_2$; and
\item $\cmatrans(q,\lab,q_2)\rightarrow q_2$.
\end{itemize}
It is easy to see that $\cma$ recognizes $\L_{dif}$.

An example of language considering both a local and a global property would be the following. Define $\L_{dif}^a$
as the language where all the positions with label $a$ have different data values. 
The transducer of a $\DA$, and the labels in the transition function of the $\CMA$ are
necessary to identify the positions labelled with $a$. 
We leave the encoding of this language as an exercise to the reader.
\end{example}

\mypar{$\FO$, $\FO^2$ and $\MSO$ over Data}
\label{sec:logicsstring}
In this section we leave the realm of automata and we concentrate instead, on declarative models like
logics.
We define \emph{First-Order Logic} (\FO) and \emph{Monadic Second-Order Logic} (\MSO) over data strings.
For a brief introduction to first and second-order logic (over data structures) see \cite{langautomlogic,David2004}.

\paragraph{First-Order Logic.}
As before, we are given an alphabet $\Ialph=\inputalph\times\dataalph$.
Given a data string $w$,
in first-order logic over strings, variables range over positions of $w$.
For every $\lab\in\inputalph$ there exists a predicate $\lab(x)$ denoting that a position $x$
contains a label $\lab$. We are also allowed to check if two variables contain the same position with the predicate $x=y$.
Given these elementary predicates we can build first-order formulae by means of logical
connectives. 
Formally formulae are defined by the following grammar:
$$\phi::=a(x)\ |\ x=y\ |\ \exists x.\phi\ |\ \phi\vee \phi\ |\ \neg\phi$$
A data string can then be seen as a model for a formula $\phi$ in this logic. 
As an example, the formula $\psi:=\forall x(a(x))$ describes the strings where all the positions are labelled with the symbol $a$.
We write $\FO$ to denote first-order logic over strings. \loris{Should I give the formal definition of the semantics?}

Let $\foelp$ be $\FO$ enriched with the following atomic binary predicates:
$x\equal y$, $x < y$, and $x=y+1$.
The predicate $x\equal y$ holds when the positions $x$ and $y$ contain the same data value,
$x<y$ holds when position $x$ occurs strictly before position $y$, and $x=y+1$ holds when $x$ is the position right
after $y$.
Given a formula $\phi$ we say that a string $w$
\emph{satisfies} $\phi$, if $w$ is a model of $\phi$. The language of $\phi$ (denoted by $\L(\phi)$)
is the set of data strings that satisfy a formula $\phi$. A formula satisfied by at least one data string is called \emph{satisfiable}
(alternatively its language is non-empty).

$\FO$ does not impose any restriction on the number of variables appearing in formulae. 
We write \FO$^v$ to denote $\FO$ with at most $v$ variables. 
The formula $\psi$ above is in \FO$^1$.
\begin{example}
\label{ex:foex1}
Consider the formula $\phi_1:=\exists x\exists y(y=x+1\wedge y\equal x)$ in $\FO^2$. 
The language $\phi_1$ is $\L_0$,
the same language accepted by the register automaton in example \ref{ex:raconsec}.
\end{example}

When we want to consider a variant of the logic in which we drop some of the predicates we just omit them from the signature.
For example $\foep$ is the same as $\foelp$ without the $<$ predicate.
The number of variables in a formula is critical for both decidability and expressiveness.
As an example consider the following case.
\begin{proposition}
$\fotel\subset \fotelp$, while, for all $v>2$, $\fokel{v}=\fokelp{v}$.
\end{proposition}
This result is due to the fact that $y=x+1$ is equivalent to the following predicate
that requires three variables:
$$x<y\wedge \forall z (x<z\implies(z=y\vee y<z))$$ 

Before moving on, we define another variant of \FO. We write $\foelw$ to denote $\FO$ with the predicates
$x\equal y$, $x < y$, and $x=y+k$ where $k\in\nat$. The first two predicates are the same as before, while
$x=y+k$ holds when $y$ is $k$ position apart from $x$ (formally, if $x$ represents the position $i$, and $y$ the position $i+k$).
Of course any given formula in this logic uses only a finite number of predicates of the form $+k$.

\paragraph{Second-Order Logic.} 
In monadic second-order logic over strings (\MSO) quantification over variables and also unary predicates (sets of variables) is allowed.
However, quantification over $\dataalph$ is not allowed.
Formally first-order logic is extended with second-order variables $X,Y,\ldots$ representing sets of positions in a string,
and the atomic predicate $x\in X$, meaning that the position in $x$ belongs to the set of positions $X$. The adjective ``monadic''
refers to the fact that we can only quantify over unary predicates and not over relations. This restriction still allows to quantify over sets, which can be 
represented as unary predicates. Formulae in $\MSO$ are generated by the following grammar:
$$\phi::=a(x)\ |\ x=y\ |\ x\in X\ |\ \exists x.\phi\ |\ \exists X.\phi\ |\ \phi\vee \phi\ |\ \neg\phi$$
We define $\msoelp$ to be $\MSO$ enriched
with the predicates over positions and data. 
The following is a classical result \cite{langautomlogic}.
\begin{proposition}
$\msoelp$ is strictly more expressive than $\foelp$.
\end{proposition}
\begin{example}
Providing an interesting example in $\MSO$ that is easy to understand is difficult. We present a simple language
that does not actually need the expressiveness of $\MSO$, however it makes $\MSO$ easier to understand.
Let's consider the formula
$$\phi_M:=\exists X(\forall x((x\in X \implies a(x))\wedge( x\notin X\implies \neg a(x) )\wedge
(x\in X\implies\forall y(y\in X\implies x\equal y))))$$
The formula describes a language $\L_M$ in which all the positions labelled with the symbol $a$ contain the same data value.
Notice that the same language can be expressed in first-order logic with the formula:
$\phi_F:=\forall x(a(x)\implies(\forall y(a(y)\implies x\equal y)))$.
\end{example}
An interesting restriction of $\MSO$ is $\EMSO$. In $\EMSO$ we only allow quantifiers over unary predicates to appear at the beginning of the formula and to be existential. 
Formally a formula $\phi$ is in $\EMSO$ if it is of the form
$\exists X_1\ldots\exists X_n (\psi)$
where $\psi$ contains no quantifiers over second-order variables.
We denote with $\emsoelp$ the corresponding logic enriched with data predicates, and with $\emsotelp$ the two-variable variant. The restriction on the number of variables only applies to first-order variables (not to the set variables).

We finally consider the logic $\emsotelt$. 
$\oplus 1$ is the successor predicate for positions belonging to the same class.
As we said in section \ref{sec:basicdef}, a \emph{class} is a maximal set of positions in a data string with the same data value (i.e. an equivalence class of the relation $\equal$). 
We have that $x=y \oplus 1$ holds iff there exists a class with positions $i_1 < \ldots < i_k$, such that
$y=i_j$ and $x=i_{j+1}$ for some $j<k$. $\oplus 1$ is called the \emph{class successor}.
Informally, the predicate holds for two positions containing the same data value $\data$ such that all the positions in between
contain elements different from $\data$.

\subsection{Expressiveness}
\label{sec:stringexpr}
\begin{figure}[htp]
\centering
\makebox[\textwidth][c]{
\input{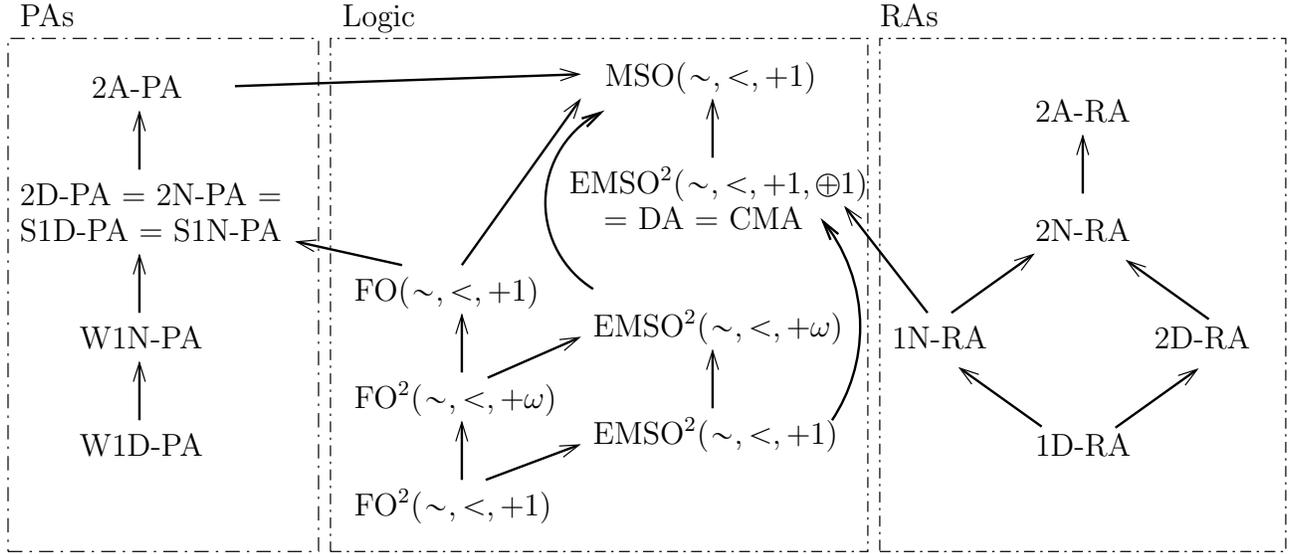}
}
\caption{Expressiveness relations. All the arrows indicate strict inclusions.}
\label{expr}
\end{figure}

As mentioned earlier, we are looking for a model that fits the notion of regularity. We presented several models,
but we cannot say anything about their properties just from their definitions. 
In this section we compare their expressiveness in terms of the set of languages they can define. The most important
expressiveness relations are summarized in figure \ref{expr}.

\mypar{Logics}
We start by summarizing the expressiveness results of the different logics we have introduced.
The first notable result is that the $<$ operator cannot be simulated in $\foep$. 
However, it can be simulated by \MSO.
\begin{proposition}
[\cite{David2004}]
The following relations hold:
\begin{itemize} 
\item $\foep \subset\foelp$,
\item $\fotep\subset\fotelp$, and
\item $\msoelp = \msoep$
\end{itemize}
\end{proposition}
Increasing the number of variables in both $\FO$ and $\MSO$ adds expressiveness. This result is interesting not only in 
terms of expressiveness. Indeed, as we will see later, both $\FO$ and $\MSO$ become undecidable when more
than two variables are allowed.
\begin{proposition}[\cite{David2004}]
For all $v\in\nat$, 
\begin{itemize}
\item $\fokelp{v} \subset\fokelp{{v+1}}$,
\item $\emsokelp{v}\subset\emsokelp{{v+1}}$, 
\item $\emsokelt{v}\subset\emsokelt{{v+1}}$, and
\item $\msokelp{v}\subset\msokelp{{v+1}}$.
\end{itemize}
\end{proposition}
We then list some results for the logics with the predicates $+\omega$ and $\oplus 1$.
While the predicate $+\omega$ does not add expressiveness in general, it cannot be expressed with the two-variable restriction.
Moreover, as expected, the predicates $\oplus 1$ and $\omega 1$ are orthogonal in expressiveness in the two-variable case.
\begin{proposition}[\cite{David2004}] The following relations hold:
\begin{itemize}
\item $\fotelp\subset\fotelw$,
\item $\emsotelp\subset\emsotelw$, and	
\item $\emsotelw\nsubseteq\emsotelt$ and $\emsotelt\nsubseteq\emsotelw$.
\end{itemize}
\end{proposition}
In general $\EMSO$ is more expressive than $\FO$. This result holds also in the two-variable case.
\begin{proposition}[\cite{David2004}] The following relations hold:
\begin{itemize}
\item $\fotelp\subset\emsotelp$, and
\item $\fotelw\subset\emsotelw$.
\end{itemize}
\end{proposition}
We now compare the previous logics and the automata models we proposed.

\mypar{Register Automata and Logic} 
The first surprising result is that deterministic register automata are in some sense \emph{orthogonal} to $\FO$ and $\MSO$
in that they cannot express properties that are definable in first-order logic, but they can express properties that are not
definable in monadic second-order logic.

\begin{proposition}[\cite{NevenFSM}]
There exists a language expressible by a {\rm{2}}\D-$\RA$ that cannot be expressed in $\msoelp$.
\end{proposition}

\noindent Since we are looking for a regular model, we would like our automata representation of data languages to have a counter part in logic.
The previous result shows that two-way register automata are probably not a ``good'' model.
One can wonder what happens with one-way $\RA$s. We have that 1\N-$\RA$s are strictly less expressive than 
\fotelt.
Things seem to get better, but unfortunately, the inclusion is strict (i.e. if the two models are equivalent).
Our last hope is that $\RA$s are at least equivalent to a weaker logic, for example $\fotelp$.
However this is not the case.
\begin{proposition}[\cite{Bojanczyk2006word}]
The language $\L_{dif}$ can be expressed in $\fotelp$ but not by a {\rm{2}}\A-$\RA$.
\end{proposition}

\noindent This result is really strong since it takes into consideration two-way alternating register automata, the most expressive model 
of $\RA$ that we have considered. Register automata do not seem to cope well with logic.
Moreover, the different variants, deterministic, non-deterministic and alternating are all increasingly more expressive:
{\rm{2}}\D-$\RA$ $\ \subset$ {\rm{2}}\N-$\RA$ $\ \subset$ {\rm{2}}\A-$\RA$. 
This result shows that the model is not determinizable.
It is worthy pointing out that the strict inclusions only hold if we assume that the complexity classes \textsc{LogSpace}, \textsc{NLogSpace} and \textsc{PTime} are all different in expressiveness.

\mypar{Pebble Automata and Logic}
After the discouraging results on register automata we now move to pebble automata. This class seems more comparable to logic than
the previous one.
$\PA$s nicely fit between first and second-order logic. 

When comparing $\PA$s to first-order logic, we have that
even the weakest form of $\PA$ is more expressive than first-order logic: $\foelp \subset$ {\rm{1}}\D-$\PA$. 
Moreover, even the strongest form of $\PA$ is included in monadic-second order logic: {\rm{2}}\A-$\PA\subset$ \msoelp.
The strictness of the containment, again, holds only under standard complexity assumptions.
Even though we have not found a logic that is exactly equivalent to some form of $\PA$, we showed that they
fit between first and second-order logic. When comparing them to logic, $\PA$s seem to behave in a \emph{regular}
way. 

The following proposition shows that $\PA$s are a robust model, in the sense that most of their variants are equivalent.
\begin{proposition}[\cite{NevenFSM}]
{\rm{2}}\D-\PA, {\rm{2}}\N-\PA, \emph{strong} {\rm{1}}\D-$\PA$ and \emph{strong} {\rm{1}}\N-$\PA$  are all equivalent models.
\end{proposition}

\noindent 
$\PA$s have appealing expressiveness properties. However, as mentioned before, a regular model
should also satisfy other properties. In this subsection we concentrated more on expressiveness. In the next one we will look at
decision problems and see that $\PA$s do not behave as well in terms of decidability. 

\mypar{Class-Memory Automata, Data Automata and Logic}
Data automata were introduced to overcome the incapability of register automata to describe \emph{local properties}. 
Therefore they increase 1$\N$-$\RA$ expressiveness while preserving decidability.
$\DA$s also have a logical counter part, \emsotelt, and are equivalent to $\CMA$s. 

As we said earlier, $\CMA$s are easier to use then $\DA$s. 
This result makes $\CMA$s ideal candidates for the role of regular data languages. 
$\CMA$ and $\DA$ are the first automata models over data strings with a logic characterization. 
Moreover, as we will see later, $\CMA$s define the `biggest' class of data languages for
which emptiness is decidable.
For completeness, we also mention that $\CMA$s are not determinizable: $\D\CMA\subset\CMA$.

We summarize the expressiveness results in Figure \ref{expr}. Some other expressiveness results are added for sake of completeness.
In the figure every arrow represents a strict inclusion.

\subsection{Decision Problems}
\label{sec:decprob}
In the this section we discuss decision problems for the models we presented.

We start from the results on $\RA$s and $\PA$s \cite{Kaminski1994,NevenFSM}.
Given a {\rm{1}}\N-$\RA$ $\ra$, checking whether a data word $w$ is accepted by $\ra$ (the membership problem) can be decided in polynomial time.
Checking whether a {\rm{1}}\N-$\RA$ does not accept any string (emptiness) is decidable. The same problems is undecidable
for 2\D-$\RA$.
Checking whether a {\rm{1}}\N-$\RA$ accepts every string (universality)
is undecidable.
The problems of inclusion and equivalence are undecidable in general. However if both the 1N-$\RA$s we are comparing have at most 2 registers checking equivalence becomes decidable.


As we mentioned in section \ref{sec:stringexpr}, the decidability results for $\PA$s are discouraging.
Indeed, even for 1\D-$\PA$ the emptiness problem and the universality problem are already undecidable. This results immediately imply that equivalence
and inclusion for 1\D-$\PA$ are also undecidable. 

Decidability results for logics are more interesting and well-understood.
\begin{proposition}[\cite{Bojanczyk2006word}]
\label{logicdec}
The satisfiability (emptiness) problem for the following logics is decidable:
\begin{itemize}
\item \fotelp,
\item \fotelw, and
\item \emsotelt.
\end{itemize}
\end{proposition}
In general satisfiability of $\foelp$ (i.e. emptiness of the underlying data language) is undecidable.
However, a clearer boundary for decidability has been established. In fact, simply allowing
$\FO$ to use three variables leads to undecidability.
This property strengthens the result of proposition \ref{logicdec}.

We finally give the precise complexity of the membership problem for $\CMA$s. 
The problem of checking whether a word $w$ is accepted by a $\CMA$ is decidable and $\comp{NP-Complete}$ \cite{Björklund07}. 

The decidability results are summarized in table 1.

\begin {table}[tp]
\label{complexitywords}
\centering
\noindent\makebox[\textwidth]{
\begin {tabular}{lllll}
\toprule 
					& 		Emptiness					& Universality 								&  Inclusion	 								&  Equivalence 							 \\\toprule

\multirow {2}*{{\rm{1}}\N-\RA} 	
				&    Dec.  								&  Dec. w/ 2 reg.						& 	Dec. w/ 2 reg.					&  Dec. w/ 2 reg.		  	 		\\
		    	&      									&  Undec.										& 	Undec.     								&  Undec.							  		 \\\midrule			
					 
{\rm{2}}\D-\RA  	&    Undec.  							&  Undec											& 	Undec.								    &  Undec.								     \\\midrule
weak {\rm{1}}\D-\PA 	&    Undec.  					  &  Undec												& 	Undec.								    &  Undec.								     \\\midrule
\fotelp      	&    Dec.  								&  Dec.											& 	?										    &  ?											 \\\midrule
\fothreeelp  &    Undec.  						&  Undec												& 	Undec.								    &  Undec									 \\\midrule
\fotelw      	&    Dec.  								&  Dec.										& 	?										  	&  ?											 \\\midrule
\emsotelp    &    Dec.  								&  ?											& 	?										   	&  ?											 \\\midrule	
\emsotelt    &    Dec.  								&  ?										& 	?										   	&  ?											 \\\bottomrule	
\end {tabular}}
\caption{Decidability results for models over data strings.}
\end{table}

\subsection{Closure Properties}
\label{sec:closureprop}
In table \ref{closprop} we present the closure properties of the models we defined.
The results are from \cite{NevenFSM,Bojanczyk2006word,Björklund07}.

\begin {table}[tp]
\label{closprop}\centering
\noindent\makebox[\textwidth]{
\begin {tabular}{lccccc}
\toprule 
					& 		$\cap$   	&  $\cup$ 		&  Concat.	 	&  Compl.   & Kleene-Star 			\\\toprule

{\rm{1}}\N-\RA      &     ?  			 &  ?					& 	?			        &  No		       & ?         					\\\midrule		 
\fotelp      	&    Yes  			&  Yes				& 	?			        &  Yes	       & ?         					\\\midrule		
\fotelw      	&    Yes  			&  Yes				&  ?					&  Yes 			& ?						 \\\midrule
\emsotelp    &    Yes  			&  Yes				& 	?					&  ?   			& ?					 \\\midrule	
\emsotelt,\DA,\CMA  &    Yes  			&  Yes				& 	Yes				&  No 			& No					 \\\bottomrule											 

\end {tabular}}
\caption{Closure properties for models over data strings. Concat. stands for concatenation and Compl. for complement.}
\end{table}
Looking at the table we can see that very few models have been carefully studied. However, whenever a model is clearly
characterized with a logic, most of the closure properties come naturally. Notice that EMSO is in general not closed under complement since
when negating a formula the existential quantifiers become universal.

\subsection{Regularity}
\label{sec:regularitystring}
In this section we discuss which models better suit the definition of regularity we gave in section 1.
None of the models we presented has all the nice properties we can find in regular string languages.
Unfortunately, we don't know if a model with all such properties even exists for data languages.
Looking at the diagrams we presented,
it is clear that the best candidates to define regular data languages are data automata and class memory automata.
We first list their positive features: 
\begin{enumerate}
\item $\DA$s and $\CMA$s are two equivalent formalisms and they have an equivalent counter part in logic, $\emsotelt$, and
\item their languages are closed under intersection, union and concatenation.
\end{enumerate}
Now we list their negative features:
\begin{enumerate}
\item the membership problem for both $\DA$s and $\CMA$s is \comp{NP-Complete},
\item $\CMA$s are not determinizable, and
\item their languages are not closed under complement and Kleene-star.
\end{enumerate}
Even though this class of languages is ``relatively small'', most of its properties are discouraging.
However there is still hope in finding a regular model. 
We have not yet found an automata model that captures $\fotelp$ or $\fotelt$. These classes are both closed under complementation.
Several extensions and restrictions of register automata have been studied \cite{Tal1999,Cheng1998}.
In particular, in \cite{Tal1999} they identified a subclass of $\RA$ with decidable equivalence.
A context-free version of register automata is proposed in \cite{Cheng1998}. This model 
adds expressiveness at the cost of slower decision procedures. However, it does not compare well with logic. 

It is clear from the results in this section (in particular, the \comp{NP} completeness of $\CMA$'s membership) that either 
a class of languages with all the regular properties does not exist, or such a class is less expressive than $\CMA$.

\section{Tree Data Languages}
\label{treedef}
String data languages allow strings to be enriched with data. However in tasks like XML reasoning, this is not enough.
Languages like XPath \cite{xpath} are able to query XML documents performing comparison on data values.
In XML documents, data appear attached to tree nodes; therefore we need a model where we can talk about 
tree structures and data at the same time.
When allowing data comparisons, XPath is known to have undecidable emptiness checking.
However, there are examples of fragments where tasks involving attribute values are decidable \cite{Arenas2005,Buneman2003,Bojanczyk2006word}.

For the automata approach most of the results are discouraging. 
\cite{Jurdzinski2008} introduces \emph{alternating register automata} over data trees. 
Emptiness is proven to be decidable for the automata that can move only down and right, and have at most one register.
Every extension in the number of registers or in the control leads to undecidability.
In \cite{Figueira2009,Figueira2011}, a \emph{bottom-up} version of such automata is introduced, and the decidability result is pushed further. 
In \cite{Bojanczyk2012} the whole XPath language is captured by an extension of data automata, which, however clearly leads to an undecidable model.

In this section we present the more interesting result for regularity of data tree languages.

\subsection{$\tfotep$ over Data Trees}
In section \ref{sec:logicsstring} we have introduced first and second-order logic over data strings.
In this section we consider $\fotep$ applied to data trees. 

\mypar{Data Trees and Logic }
We consider unranked, ordered, labelled trees with data values.
A \emph{data tree} $t$ over $\Ialph=\inputalph\times\dataalph$ is a tree where every node is a datum $(\lab,\data)\in\Ialph$.
In the same way as for data strings, a data tree can be seen as a model for a logic formula. 
We consider \FO. 
Now variables range over nodes instead of position.
Similar to the case of data strings we will have the following predicates.
For every $\lab\in\inputalph$ there is a predicate $\lab(x)$ denoting that a node $x$
contains a label $a$.
Let $\tfotelp$ be two-variable first-order logic over trees, enriched with the following atomic predicates:
$x\equal y$, $E_{\rightarrow}(x,y)$, $E_{\downarrow}(x,y)$, $E_{\rightarrow}\st(x,y)$, and $E_{\downarrow}\st(x,y)$.
The binary predicate $x\equal y$ holds when the nodes $x$ and $y$ contain the same data value.
The binary predicate $E_{\rightarrow}(x,y)$ holds if $x$ and $y$ have the same parent and $y$ is the immediate sibling
of $x$ (the next child in order). 
$E_{\downarrow}(x,y)$ holds if $y$ is a child of $x$.
$E_{\rightarrow}\st(x,y)$, and $E_{\downarrow}\st(x,y)$ are the reflexive transitive closures of $E_{\rightarrow}(x,y)$, and $E_{\downarrow}(x,y)$ respectively.
We write $\tfotep$ to indicate the same logic without the last two predicates.

\mypar{Expressiveness and Decidability}
The main decidability result on data trees is the following.
\begin{proposition}[\cite{Bojanczyk2006tree}]
\label{treeemptiness}
The logic $\tfotep$  on unranked trees is decidable in \comp{3NExpTime}. 
\end{proposition}
This is the biggest class of tree data languages supported by a logic for which emptiness is decidable.
However this bound is not proven.
In the same paper, the more expressive logic $\tfotelp$ is also analysed. 
In the case of data strings, the extended logic $\fotelp$ is still decidable. 
When we have data trees, emptiness for this logic is proven to be a `hard' open problem.
In fact, the problem of emptiness for vector addition tree automata \cite{Groote2004} is reduced to satisfiability for $\tfotelp$. 
Emptiness for vector addition tree automata is equivalent to several notorious open problems. 
Therefore proving decidability for $\tfotelp$ is challenging.
To understand better why this is a difficult problem we give some intuition of how hard the proof of proposition \ref{treeemptiness} is.

\noindent\emph{Proof Sketch. }
The proof \cite{Bojanczyk2006tree} proceeds as follows:
1) first, the satisfiability problem for $\tfotep$ on unranked data trees is reduced to a \emph{puzzle} problem that asks whether a data
tree with some properties exists, then
2) a pumping argument on the size of solutions of such puzzle is given, finally,
3) a procedure for looking for such solutions is given using a class of tree automata with decidable emptiness checking.

We start discussing the first step. 
We do not give the formal definition of the puzzles. However, 
a puzzle can be seen as a variation of a data automaton over data trees.
Solving the puzzle means finding a data tree that has a run over the automaton and has
some restrictions on how many times each label in $\inputalph$ can appear in every data class.
A formula in $\tfotep$ can be converted to an instance of the puzzle as follows:
a) the original formula in $\tfotep$ is converted into an equivalent formula in $\temsotep$ in ``data normal form'' (with some restrictions),
b) from a formula in ``data normal form'' we then compute a puzzle that has solutions iff the formula is satisfiable.
The formula is satisfiable iff the puzzle has a solution.

The second step of the proof, the pumping argument, is the hardest.
In this step it is shown that if a puzzle admits a solution, it also admits a solution of bounded size. 
Moreover this size is effectively computable.
This steps uses the notion of zone. In a data tree, a \emph{zone} is a maximal connected set of nodes with the same data value.
Given a solution of the puzzle, the solution can be transformed into one where only $M$ zones contain more than $N$ nodes. 
$M$ and $N$ are effectively computable and they only depend on the size of the puzzle. We call the transformed solution an $(M,N)$-reduced solution. 
The reduction proceeds by steps. Each step is rather technical but follows the same pattern:
zones with `problematic' properties are identified and they are transferred into other zones by changing their data values.
This modification preservers the property of being a solution but decreases the number of `problematic' zones.

The third and final step is that of finding whether a given puzzle has a $(M,N)$-reduced solution.
The algorithm is by reduction to the emptiness of linear constraint tree automata. 
\emph{Linear constraint tree automata} (LCTA) is a tree automata $T$ together with a linear constraint on the set of states of $T$.
A run of an LCTA is accepting if, when we instantiate the constraint with the number of times each state appears in the run, the constraint
is satisfied. 
Given a data tree we call the tree with labels and no data its data erasure.
Given a puzzle $P$ and numbers $M,N$, one can compute an LCTA that recognizes the data erasure of the $(M,N)$-reduced solutions of $P$.
The nice idea of this step is that of reducing a problem over data trees to a problem where no data is considered and therefore using known models.

The complexity obtained this way is \comp{3NExpTime}
which is possibly not optimal.

\subsection{Regularity} 
\label{sec:regularitytrees}
The only logic for which  satisfiability was proved to be decidable is $\tfotep$. 
However, it is too early to talk about regularity, since we do not know any automata model equivalent to $\tfotep$.
Indeed regularity of data tree languages is a real open problem.

Several extensions of $\RA$s, $\PA$s, and $\DA$s have been proposed to deal with data trees \cite{Cheng1998,Kaminski2008,Bojanczyk2012},
even though none of these extensions addresses the problem of regularity.
In \cite{Kaminski2008}, a complex extension of $\RA$s for trees was introduced, called \emph{unification-based automata} (UBA).
Emptiness is decidable for UBA, and the class is robust in the sense that the expressiveness of the bottom-up and top-down versions are 
equivalent. However the same expressiveness restriction of the $\RA$s over the string case applies.

In \cite{Cheng1998} a pushdown version of register automata has been defined (PDRA). This extension
can be used for both string and tree data languages. Emptiness for PDRA is decidable, 
but the model does not seem to fit in any logic characterization.
In \cite{Bojanczyk2012}, a complex extension of class-memory automata over data trees is defined. This class
has a clear logic counterpart and good closure properties. Unfortunately, the emptiness problem for $\CMA$s over data trees is undecidable, which makes them unlikely to be a good regular model. This class is, however, useful for practical purposes, since it captures a specific fragment of XPath. 





\section{Conclusions and Open Problems}
\label{discussion}
We presented a variety of models for representing data languages, both for data strings and data trees.
Several others have been proposed in literature \cite{Bouyer2003}, 
but we focused our attention on the historically interesting models
in the area of regularity of data languages.
We started with register and pebble automata, two classic models that were adapted to the data setting.
We then explored new ideas such as data automata, class-memory automata and their corresponding logics.
Even though the situation is much less favourable than for regular string languages, $\CMA$s have some desirable properties.
When comparing them to the other proposed models, $\CMA$s appear to have a better trade-off between decidability and 
expressiveness.

Then we described how logic and data automata do not extend smoothly to data trees. In this area, the state of the art of
regularity is discouraging. The few decidable models are not expressive enough, and automata and logic seem to travel on
different tracks. Even for simple classes of tree data languages most of the decision problems are still open, and very few usable
models have been proposed and analysed.

The field of regular data languages is still open, and the landscape is not clearly understood. We still need to identify the ``right'' classes
of data languages. Here, we list some interesting open problems related to regularity:
\begin{itemize}
\item Is there a variant of $\RA$ (section \ref{sec:regaut}) closed under complementation and with no restriction on the number of registers?
\item Is there a version of $\FO$ (section \ref{sec:logicsstring}) equivalent to one of the variants of $\RA$?
\item Is there a logic that is closed under complementation and has decidable equivalence?
\item Is satisfiability of $\emsotelw$ and $\emsoteltw$ (section \ref{sec:logicsstring}) decidable?
\item Is there an automata model corresponding to $\emsotelw$ (section \ref{sec:logicsstring})?
\item Is satisfiability of $\tfotelp$ (section \ref{treedef}) decidable?
\item Is there a variant of $\CMA$ (section \ref{treedef}) over trees with decidable emptiness?
\end{itemize}

In this paper, we focused our attention on data strings and data trees. However, several results have been extended to more complex models
such as \emph{infinite data strings} \cite{Bojanczyk2006word}.
These extensions are very useful in verification and model checking. One family of models we have not considered, that easily adapts
to this setting, is that of temporal logics over data strings \cite{Demri2009}. 
These models have good decision procedures but
are not relevant for regularity. 

Finally, the field of transducers lies completely open for development.
The idea of \emph{symbolic transducers} proposed in \cite{Veanes2012}
is, to our knowledge, the only variant of transducers over infinite alphabets. The framework proposed in \cite{Veanes2012}
is, however, different from the one we considered.
In fact, the set of operations over the data is not restricted to equality. Instead, the set of operations depends
on the domain considered.
No model has been proposed for \emph{data string transducers} (and \emph{data tree transducers})
and there is still work to be done in the area.

\bibliographystyle{alpha}
\bibliography{refs}

\newcommand{\etalchar}[1]{$^{#1}$}
\begin{thebibliography}{AMN{\etalchar{+}}01}

\bibitem[AC11]{SSTAlur}
Rajeev Alur and Pavol Cern\'{y}.
\newblock Streaming transducers for algorithmic verification of single-pass
  list-processing programs.
\newblock {\em SIGPLAN Not.}, 46(1):599--610, January 2011.

\bibitem[AFL05]{Arenas2005}
Marcelo Arenas, Wenfei Fan, and Leonid Libkin.
\newblock On verifying consistency of xml specifications, 2005.

\bibitem[AMN{\etalchar{+}}01]{Alon2001}
Noga Alon, Tova Milo, Frank Neven, Dan Suciu, and Victor Vianu.
\newblock Xml with data values: typechecking revisited.
\newblock In {\em Proceedings of the twentieth ACM SIGMOD-SIGACT-SIGART
  symposium on Principles of database systems}, PODS '01, pages 138--149, New
  York, NY, USA, 2001. ACM.

\bibitem[BDF{\etalchar{+}}03]{Buneman2003}
Peter Buneman, Susan Davidson, Wenfei Fan, Carmem Hara, and Wang-Chiew Tan.
\newblock Reasoning about keys for xml, 2003.

\bibitem[BDM{\etalchar{+}}06]{Bojanczyk2006tree}
Mikolaj Boja\'{n}czyk, Claire David, Anca Muscholl, Thomas Schwentick, and Luc
  Segoufin.
\newblock Two-variable logic on data trees and xml reasoning.
\newblock In {\em Proceedings of the twenty-fifth ACM SIGMOD-SIGACT-SIGART
  symposium on Principles of database systems}, PODS '06, pages 10--19, New
  York, NY, USA, 2006. ACM.

\bibitem[BHM02]{Bouajjani2002}
Ahmed Bouajjani, Peter Habermehl, and Richard Mayr.
\newblock Automatic verification of recursive procedures with one integer
  parameter, 2002.

\bibitem[BL12]{Bojanczyk2012}
Mikolaj Bojanczyk and Slawomir Lasota.
\newblock An extension of data automata that captures xpath.
\newblock {\em CoRR}, abs/1201.0597, 2012.

\bibitem[BMS{\etalchar{+}}06]{Bojanczyk2006word}
Mikolaj Bojanczyk, Anca Muscholl, Thomas Schwentick, Luc Segoufin, and Claire
  David.
\newblock Two-variable logic on words with data.
\newblock In {\em Proceedings of the 21st Annual IEEE Symposium on Logic in
  Computer Science}, LICS '06, pages 7--16, Washington, DC, USA, 2006. IEEE
  Computer Society.

\bibitem[BPT03]{Bouyer2003}
Patricia Bouyer, Antoine Petit, and Denis Thérien.
\newblock An algebraic approach to data languages and timed languages, 2003.

\bibitem[BS07]{Björklund07}
Henrik Björklund and Thomas Schwentick.
\newblock On notions of regularity for data languages.
\newblock In {\em In FCT}, pages 88--99, 2007.

\bibitem[CK98]{Cheng1998}
Edward~Y.C. Cheng and Michael Kaminski.
\newblock Context-free languages over infinite alphabets.
\newblock {\em Acta Informatica}, 35:245--267, 1998.
\newblock 10.1007/s002360050120.

\bibitem[Dav04]{David2004}
Claire David.
\newblock Mots et donne\'{e}s infinis. master's thesis, 2004.

\bibitem[DL09]{Demri2009}
St{\'e}phane Demri and Ranko Lazi\'{c}.
\newblock Ltl with the freeze quantifier and register automata.
\newblock {\em ACM Trans. Comput. Logic}, 10(3):16:1--16:30, April 2009.

\bibitem[DSV99]{xpath}
Alin Deutsch, Liying Sui, and Victor Vianu.
\newblock Xml path language (xpath) version 1.0. w3c recommendation, the world
  wide web consortium.
\newblock In {\em Journal of Computer and System Sciences (JCSS) 2007;
  73(3):442–474}, 1999.

\bibitem[Fig09]{Figueira2009}
Diego Figueira.
\newblock Satisfiability of downward xpath with data equality tests.
\newblock In {\em Proceedings of the twenty-eighth ACM SIGMOD-SIGACT-SIGART
  symposium on Principles of database systems}, PODS '09, pages 197--206, New
  York, NY, USA, 2009. ACM.

\bibitem[FS11]{Figueira2011}
Diego Figueira and Luc Segoufin.
\newblock Bottom-up automata on data trees and vertical xpath.
\newblock In {\em STACS}, pages 93--104, 2011.

\bibitem[GGS04]{Groote2004}
Philippe~de Groote, Bruno Guillaume, and Sylvain Salvati.
\newblock Vector addition tree automata.
\newblock In {\em Proceedings of the 19th Annual IEEE Symposium on Logic in
  Computer Science}, LICS '04, pages 64--73, Washington, DC, USA, 2004. IEEE
  Computer Society.

\bibitem[HMP92]{Henzinger92timedtransition}
Thomas Henzinger, Zohar Manna, and Amir Pnueli.
\newblock Timed transition systems, 1992.

\bibitem[JL08]{Jurdzinski2008}
Marcin Jurdzinski and Ranko Lazic.
\newblock Alternating automata on data trees and xpath satisfiability.
\newblock {\em CoRR}, abs/0805.0330, 2008.

\bibitem[KF94]{Kaminski1994}
Michael Kaminski and Nissim Francez.
\newblock Finite-memory automata.
\newblock {\em Theor. Comput. Sci.}, 134(2):329--363, November 1994.

\bibitem[KT08]{Kaminski2008}
Michael Kaminski and Tony Tan.
\newblock Pillars of computer science.
\newblock chapter Tree automata over infinite alphabets, pages 386--423.
  Springer-Verlag, Berlin, Heidelberg, 2008.

\bibitem[NSV04]{NevenFSM}
Frank Neven, Thomas Schwentick, and Victor Vianu.
\newblock Finite state machines for strings over infinite alphabets.
\newblock {\em ACM Trans. Comput. Logic}, 5(3):403--435, July 2004.

\bibitem[Seg06]{Segoufin2006}
Luc Segoufin.
\newblock Automata and logics for words and trees over an infinite alphabet.
\newblock In {\em Proceedings of the 20th international conference on Computer
  Science Logic}, CSL'06, pages 41--57, Berlin, Heidelberg, 2006.
  Springer-Verlag.

\bibitem[Tal04]{Tal1999}
A.~Tal.
\newblock Decidability for inclusion for unification based automata, 2004.

\bibitem[Tho96]{langautomlogic}
Wolfgang Thomas.
\newblock Languages, automata, and logic.
\newblock In {\em Handbook of Formal Languages}, pages 389--455. Springer,
  1996.

\bibitem[VHL{\etalchar{+}}12]{Veanes2012}
Margus Veanes, Pieter Hooimeijer, Benjamin Livshits, David Molnar, and Nikolaj
  Bjorner.
\newblock Symbolic finite state transducers: algorithms and applications.
\newblock In {\em Proceedings of the 39th annual ACM SIGPLAN-SIGACT symposium
  on Principles of programming languages}, POPL '12, pages 137--150, New York,
  NY, USA, 2012. ACM.

\end{thebibliography}

\end{document}